\documentstyle[12pt]{article}

\begin{document}

\newcommand{\bda}{\begin{displaymath}\begin{array}{rl}}
\newcommand{\eda}{\end{array}\end{displaymath}}
\newcommand{\be}{\begin{equation}}
\newcommand{\ee}{\end{equation}}
\newcommand{\bea}{\begin{eqnarray}}
\newcommand{\eea}{\end{eqnarray}}
\newcommand{\bdm}{\begin{displaymath}}
\newcommand{\edm}{\end{displaymath}}
\newcommand{\no}{\nonumber \\}
\newcommand{\Mone}{M(s,t,u)_{\mbox{\tiny one loop}}}
\newcommand{\Mbar}{\;\overline{\rule[0.75em]{0.8em}{0em}}\hspace{-1em}M}
\newcommand{\MI}{M_{\mbox{\scriptsize I}}}
\newcommand{\MIhat}{\hat{M}_{\mbox{\scriptsize I}}}
\newcommand{\I}{{\mbox{\scriptsize I}}}
\newcommand{\Ism}{{\mbox{\tiny I}}}
\newcommand{\mbar}{\overline{\rule[0.5em]{0.7em}{0em}}\hspace{-0.75em}m}
\newcommand{\Dbar}{\;\overline{\rule[0.75em]{0.6em}{0em}}\hspace{-0.65em}\Delta}
\newcommand{\ubar}{\overline{\rule[0.42em]{0.4em}{0em}}\hspace{-0.5em}u}
\newcommand{\dbar}{\,\overline{\rule[0.7em]{0.4em}{0em}}\hspace{-0.6em}d}
\newcommand{\sbar}{\overline{\rule[0.45em]{0.4em}{0em}}\hspace{-0.5em}s}
\newcommand{\cbar}{\overline{\rule[0.42em]{0.4em}{0em}}\hspace{-0.5em}c}
\newcommand{\abar}{\overline{\rule[0.42em]{0.4em}{0em}}\hspace{-0.5em}a}
\newcommand{\bbar}{\,\overline{\rule[0.65em]{0.4em}{0em}}\hspace{-0.45em}b}
\newcommand{\qbar}{\overline{\rule[0.42em]{0.4em}{0em}}\hspace{-0.5em}q}
\newcommand{\Wbar}{\;\overline{\rule[0.75em]{1em}{0em}}\hspace{-1em}W}
\newcommand{\Kbar}{\,\overline{\rule[0.75em]{0.7em}{0em}}\hspace{-0.85em}K}
\newcommand{\Lbar}{\,\overline{\rule[0.7em]{0.5em}{0em}}\hspace{-0.75em}{\cal
L}}
\newcommand{\eff}{{e\hspace{-0.1em}f\hspace{-0.18em}f}}
\newcommand{\QCD}{{\mbox{\scriptsize Q\hspace{-0.1em}CD}}}
\newcommand{\sA}{s_{\hspace{-0.15em}A}}
\newcommand{\DGMO}{\Delta_{\mbox{\tiny GMO}}}
\newcommand{\mixingangle}{\theta_{\eta^\prime\hspace{-0.05em}\eta}}
\newcommand{\R}{{\scriptscriptstyle R}}
\renewcommand{\L}{{\scriptscriptstyle L}}
\newcommand{\al}{&\!\!\!\!}
\newcommand{\fs}{\; \; .}
\newcommand{\co}{\; \; ,}
\newcommand{\QD}{Q_{\hspace{-0.1em}D}}
\newcommand{\sqg}{\raisebox{0.35em}{$\sqrt{\rule[-0.2em]{1em}{0em}}$}
\hspace{-0.9em}\rule{0.03em}{0.6em}\hspace{0.1em}g
\hspace{0.1em}\rule{0.03em}{0.6em}}
\newcommand{\sqH}{\raisebox{0.35em}{$\sqrt{\rule[-0.1em]{3.1em}{0em}}$}
\hspace{-3em}\rule{0.03em}{0.7em}\det H
\hspace{0.1em}\rule{0.03em}{0.7em}}
\newcommand{\Hbar}{\;\overline{\rule[0.75em]{0.7em}{0em}}\hspace{-0.9em}H}
\newcommand{\PB}{\hspace{-0.1em}\raisebox{-0.5em}{\scriptsize PB}}

\begin{titlepage}
\rule{0em}{2em}\vspace{2em}
\begin{center}

{\LARGE {\bf Phonons as Goldstone bosons}}
\\ \vspace{0.8cm}
H. Leutwyler\\Institut f\"{u}r theoretische Physik der Universit\"{a}t
Bern\\Sidlerstr. 5, CH-3012 Berne, Switzerland and\\
CERN, CH-1211 Geneva, Switzerland\\
\vspace{2em}
June 1996\\
\vspace{3em}
{\bf Abstract} \\
\vspace{2em}
\parbox{30em}{The implications of the hidden, spontaneously broken symmetry
for the properties of the sound waves of a solid are analyzed. Although the
discussion does not go beyond standard wisdom, it presents
some of the known results from a different perspective. In particular, I
argue that, as
a consequence of the hidden symmetry, the equations of motion for
a sound wave necessarily
contain nonlinear terms, describing phonon-phonon-scattering and emphasize
the analogy with the low energy theorems valid for $\pi\pi$-scattering.}\\
\vspace{4em}\noindent  {\it Dedicated to Klaus Hepp and Walter Hunziker}

\vspace{5em}
\rule{30em}{.02em}\\
{\footnotesize Work
supported in part by Schweizerischer Nationalfonds}
\end{center}
\end{titlepage}

\section{Introduction}

Solids represent configurations that spontaneously break
translation invariance \cite{Guralnik,Lange,Nielsen}. The purpose of the
present paper
is to discuss some of the consequences of this fact, extending the analysis
given in ref. \cite{Nonrel}. Although none
of the statements made below goes beyond what is known about the physics of a
solid, it is of interest to view the low energy structure
from the point of view of general effective field theory. As I will
point out, some of the properties of the sound waves, in particular the fact
that phonons necessarily interact with one another, are intimately related to
the structure of the spontaneously broken symmetry group.

The hidden symmetry manifests itself through the occurrence of conserved
currents. In the present case, the relevant charges are the generators
of the translations, i.e. the total momentum $P^a$. The corresponding
currents are the components $\theta^{\mu a}$ of the energy-momentum-tensor,
\bdm P^a=\int\!\!d^3\!x\,\theta^{0a}(x)\fs\edm
In local form, the corresponding conservation law reads
\bdm\partial_0\theta^{0a}(x)+
\partial_b\theta^{ba}(x)=0\co\edm
where $\theta^{ab}$ is the stress tensor.
The analogue of current algebra is the commutation
relation
\bdm [\theta^{0a}(x),\theta^{0b}(x)]\raisebox{-0.5em}{$\scriptstyle \;x^0=y^0$}
=-i\hbar\partial_a
\delta^3(x-y)\theta^{0b}(x)-i\hbar\partial_b\delta^3(x-y)\theta^{0a}(y)\fs
\edm
In the following, current algebra plays an important role, but I do not need
it in this local form. Instead, I only make use
of the commutation relation
\bdm [P^a,\theta^{\mu\nu}(x)]=\hbar i\partial_a\theta^{\mu\nu}(x)\co\edm
which expresses the fact that the momentum generates
translations.

Translation
symmetry is peculiar in that, up to a factor of $c^2$, the time components
of the corresponding currents, $\theta^{0a}=\theta^{a0}$,
at the same time also represent the energy flow
and thus occur in the energy conservation law\footnote{
In the notation used here, the time variable is denoted
by $x^0\!=\!t$ (no factor of $c$). Accordingly, the energy density is given by
$c^2\theta^{00}$.}
\bdm\partial_0\theta^{00}(x)+\partial_a\theta^{a0}(x)=0\fs\edm
The operator of total energy is the Hamiltonian,
\bdm H=\int\!\!d^3\!x\,c^2\theta^{00}(x)\co\edm
which generates translations in the time direction,
\bdm [H,\theta^{\mu\nu}(x)]=-\hbar i\partial_0\theta^{\mu\nu}(x)\fs\edm

The spontaneous breakdown of
a symmetry implies that the spectrum of the Hamiltonian does not
contain an energy gap. In the present case, where the symmetry group of
interest is the translation group, the Goldstone bosons associated with
the spontaneous symmetry breakdown are the phonons and the Goldstone theorem
reduces to the well-known statement that phonons of sufficiently large
wavelength carry arbitrarily little energy.
There may be other degrees of freedom without an energy gap. In the case of a
conductor, for instance, the excitation of electrons near the fermi
surface also requires arbitrarily little energy.
For a general discussion of effective field theories related to the
electronic degrees of freedom, I refer to \cite{Froehlich}.
In the following, I concentrate on the phonons (more precisely,
on the acoustic branch of the dispersion curve) and study their behaviour at
large wavelength:
$\lambda\!\gg\! a$, where $a$ is the lattice spacing.

The low energy properties of the Goldstone bosons have been analyzed in
considerable detail for the case where the spontaneous symmetry breakdown
preserves Lorentz invariance (see \cite{Weinberg 1979,GL 1984/85,Foundations}
and the references therein). In particular, it is known that the
Goldstone bosons generated by the spontaneous
breakdown of an exact nonabelian symmetry are subject to an interaction that
grows with the square
of the momentum, the strength being determined by the transition
matrix elements of the currents between the vacuum and states containing
one Goldstone particle. In the case of an abelian symmetry group, on the other
hand, symmetry implies that the coefficient of the term of order $p^2$
vanishes, so that the Taylor series expansion of the elastic scattering
amplitude only starts at order $p^4$. Since the translations form an
abelian group, one might expect that the phonon-phonon-interaction belongs to
this second category.
The general analysis mentioned above does, however,
not apply here, because it makes essential use of the assumption that the
ground state is Lorentz
invariant, which drastically simplifies the structure of the effective
theory. More importantly,
the form of the Ward identities is controlled by the {\it local} version of the
symmetry group, not by the global one \cite{Nonrel}. The local form of the
translation group
is the set of general coordinate transformations and is not abelian.
The intrinsic difference in the structure of the global and local
versions of the symmetry group shows up in the current algebra
commutation rules: while the generators $P^a$ of the translation
group commute among themselves, the momentum densities do not.
The case under consideration thus corresponds to a nonabelian
local symmetry. As we will see, the fact that the commutator of two
momentum densities does not vanish requires
the phonons to interact among themselves.

\section{Effective Lagrangian}
The consequences of the hidden, spontaneously broken symmetry may be
analyzed in terms of an effective field theory.
The relevant effective field is a space-time-dependent element of
the translation group, which I denote by
$\xi_a(x)\!=\!\xi_a(t,\vec{x}\,),\,a\!=1,2,3$. The effective field represents
the displacement of
the material from the position in the ground state. The corresponding effective
Lagrangian ${\cal L}_\eff={\cal
L}_\eff(\xi,\dot{\xi},\partial\xi,\ddot{\xi},\partial\dot{\xi},
\partial\partial\xi, \ldots ) $ may be analyzed by means of an expansion in
powers of $\xi$:
\bdm {\cal L}_\eff={\cal L}_0+{\cal L}_1+{\cal L}_2+{\cal L}_3+
\ldots \edm
The first term is an irrelevant constant, while all others
contain
a string of contributions with an increasing number of derivatives. The
derivative expansion of ${\cal L}_2$, for example, starts with:
\bdm
{\cal L}_2= a^1_{ab}\xi_a\xi_b+a^2_{ab}\xi_a\dot{\xi}_b+
a^3_{abc}\xi_a\partial_b\xi_c+
a^4_{ab}\dot{\xi}_a\dot{\xi}_b+a^5_{abc}\dot{\xi}_a\partial_b\xi_c
+a^6_{abcd}\partial_a\xi_b\partial_c\xi_d+
\ldots\edm
The expression accounts for all terms with less than three derivatives
(contributions of the type $\xi\!\times\!\ddot{\xi}$ and
$\xi\!\times\!\partial\partial\xi$ may be eliminated with an integration by
parts).
The conservation laws and commutation relations associated with the hidden
symmetry strongly constrain
the coefficients $a_{ab}^1,a_{ab}^2\ldots$ In
fact, the properties of the effective field theory that describes the low
energy structure in a model-independent way follow from
these constraints.

The lattice of a solid does not
possess any continuous symmetries, except for the translations in the
direction of the time axis. Accordingly, the various
coupling constants occurring in the effective
Lagrangian for the phonons are only restricted by discrete symmetries, such as
reflections on lattice planes \cite{Landau Lifshitz}.
I assume that the system is invariant under space reflections and
time reversal. This implies that time derivatives only enter pairwise and
that
the Lagrangian only contains terms for which the number of space derivatives
plus the number of fields is even.
The quantity ${\cal L}_1$ then exclusively involves total derivatives and may
thus be dropped. In the case of ${\cal L}_2$, reflection
symmetry implies that the coefficients $a^2_{ab},a^3_{abc}$
and $a^5_{abc}$ vanish, etc.

For a solid with
cubic symmetry, the terms $a^1_{ab}\,\xi_a\xi_b$ and
$a^4_{ab}\,\dot{\xi}_a\dot{\xi}_b$ are of the form $a_1\,\xi_a\xi_a$ and
$a_2\,\dot{\xi}_a\dot{\xi}_a$, respectively, so that, for these contributions,
reflection
symmetry implies invariance under rotations. In the case of the coefficient
$a^6_{abcd}$, cubic symmetry permits the following terms:
\bdm a^6_{abcd}\partial_a\xi_b\partial_c\xi_d=
a_3\,\partial_a\xi_a\partial_b\xi_b +a_4\partial_a\xi_b\partial_a\xi_b
+a_5 \partial_a\xi_b\partial_b\xi_a+a_6\sum_a\partial_a\xi_a\partial_a\xi_a
\fs \edm
Only the first three are invariant under rotations. In the following,
I simplify the bookkeeping by restricting myself to those terms in the
effective Lagrangian that are invariant under rotations, thus ignoring the
term $\propto\!a_6$. The same line of
reasoning should go through also in the general case, but I did not carry
it out. Hence the following discussion does not immediately apply to a
real solid, but concerns a situation where the ground state spontaneously
breaks translation symmetry while preserving rotation invariance, as for
fluids or gases.

With an integration by parts, the term $\propto\!a_5$ may be absorbed in $a_3$,
so that only two effective couplings remain, which represent
the torsion and compression modules of the
system, denoted by $\mu$ and $K$, respectively. Also,
with a suitable normalization of the field $\xi$, the coefficient of the term
$\dot{\xi}^a\dot{\xi}^a$ may be identified with the mass density $\rho_0$,
so that ${\cal L}_2$ takes the form
\bdm {\cal
L}_2=\mbox{$\frac{1}{2}$}\rho_0\, \dot{\xi}_a\dot{\xi}_a
               -\mbox{$\frac{1}{2}$}\mu\,\partial_a\xi_b\partial_a\xi_b
               -\mbox{$\frac{1}{6}$}(\mu+3K)\partial_a\xi_a\partial_b\xi_b
             +l_0\,\xi_a\xi_a +O(p^4)
\edm
The general expression up to and including two
derivatives thus agrees with the standard form of the Lagrangian describing
the sound waves, except for the term $l_0\,\xi_a\xi_a$, which does not
occur in the standard analysis.

\section{Energy-momentum-tensor}
Indeed, the conservation laws that follow from the
presence of a hidden symmetry imply $l_0\!=\!0$.
To verify this statement, consider the
representation of the energy-momentum-tensor in terms of
the effective field,
\bdm\theta^{\mu\nu}=\theta^{\mu\nu}_0
+\theta^{\mu\nu}_1+\theta^{\mu\nu}_2+O(\xi^3 )\fs \edm
The field independent term $\theta^{\mu\nu}_0$ represents the energy
density and the pressure in the ground state. Since the derivatives thereof
vanish, this contribution is conserved by itself.
The expansion of the momentum density starts with the familiar
term proportional to the velocity:
\bdm \theta^{0a}_1=\rho_0\,\dot{\xi}_a +O(p^3)\fs\edm
Energy conservation then requires that the energy density contains a
corresponding contribution proportional to the divergence of the effective
field,
\bdm \theta^{00}_1=-\rho_0 \partial_a\xi_a+O(p^3)\co\edm
which describes the change in the rest energy caused by the deformation
$\xi_a(x)$. To first order in $\xi$, the general expression for the
stress-tensor is of the form
\bdm \theta^{ab}_1=k_1(\partial_a\xi_b+\partial_b\xi_a)+k_2\delta_{ab}
\partial_c\xi_c +O(p^3)\co\edm
and momentum conservation requires
\bdm \rho_0\,\ddot{\xi}_a+k_1\partial_b\partial_b\xi_a
+(k_1+k_2)\partial_a\partial_b\xi_b
=O(\xi^2,p^4)\fs\edm
This is consistent with the equation of motion that follows from the effective
Lagrangian of the preceding section if and only if
\bdm l_0=0\co\;\;\; k_1=-\mu\co\;\;\;k_2=\mbox{$\frac{2}{3}$}\mu-K\fs\edm
As claimed above, the coefficient $l_0$ thus vanishes.

It is clear, however, that these contributions do not represent the full
energy-momentum tensor. In particular, the energy of the sound waves
must give rise to a contribution in $\theta^{00}$ that is quadratic in
the field $\xi_a(x)$.
In his diploma work \cite{Willers}, Moritz Willers investigated the
energy-momentum-tensor belonging to the
Lagrangian ${\cal L}_2$, using the Noether theorem and run into difficulties
rather immediately: although Noether's construction yields a conserved
energy-momentum-tensor $\theta^{\mu\nu}_N$, the result fails
to be symmetric under an interchange of $\mu$ and $\nu$. The conservation
laws fix the form of the energy-momentum-tensor only up to
\bdm \theta^{\mu\nu}=\theta^{\mu\nu}_N+\partial_\lambda \chi^{\lambda\mu\nu}\co
\edm
with $\chi^{\lambda\mu\nu}\!=\!-\chi^{\mu\lambda\nu}$.
Indeed we may exploit this freedom
to arrive at a symmetric tensor, but
only at a price: the relevant expression for $\chi^{\lambda\mu\nu}$
contains the field $\xi_a$
itself, not only its derivatives -- unless $\mu\!=\!\rho_0\,
c^2,\,K\!=\!-\frac{1}{3}\rho_0\,
c^2$, in which case sound propagates with the velocity of light.
For realistic values of the torsion and compression modules, the
energy-momentum-tensor thus fails to be translation invariant.

The symmetry
requirement $\theta^{\mu\nu}=\theta^{\nu\mu}$ is perfectly physical: the
energy-momentum-tensor represents the source of gravity and can only enter
Einstein's equations if it is symmetric. Also, the
expectation that the result should be translation invariant is well
justified -- otherwise the distribution of energy and momentum depends on the
absolute position of the body. The only way out is to conclude that the
Lagrangian ${\cal L}_2$ cannot be the full story -- the equation
that describes the propagation of sound must contain nonlinear terms.

\section{Covariant formulation}
In the standard analysis of elastic deformations, the phenomenon mani\-fests
itself as follows. It is convenient not to work with a
set of three variables specifying the displacement of the various
points, but to use three scalar fields $z_a(x)$ with the property that the
world lines of the body-fixed points are characterized by
$z_a(x)\!=\!\mbox{constant}$.
The variables $\xi_a(x)$ used previously may be converted into this language by
setting
\bdm z_a(t,\vec{x}\,)=x_a-\xi_a(t,\vec{x}\,)\fs\edm
To first order in the deformation,
the body-fixed point with coordinate $z_a$ is then described by the world line
$x_a(t)=z_a+\xi_a(t,\vec{z}\,)+\ldots\,$, so that $\xi_a(x)$ indeed represents
the displacement vector. Note that, in the framework used here, the precise
physical significance of the variable $\xi_a(x)$ is left open --
with the above interpretation, $\xi_a(x)$ differs from the
displacement vector at higher orders of the deformation.

This formalism readily lends itself
to a generally covariant formulation of the dynamics. We may consider a
curved space-time with metric $g_{\mu\nu}(x)$ and form the
covariant derivatives
\bdm \partial_\mu z_a\co\;\;\;\;\nabla_\mu\partial_\nu z_a=
\partial_\mu\partial_\nu z_a-\Gamma^\lambda_{\mu\nu}\partial_\lambda
z_a\co\;\;\ldots\edm
If the function $F(\partial z,\nabla\partial
z,\ldots)_g$ is a scalar with respect to general coordinate transformations,
then the equations of motion that follow from the Lagrangian
\bdm {\cal L}_\eff=\sqg\;F(\partial z,\nabla\partial
z,\ldots)_g\fs\edm
automatically ensure that the
variational derivative of the
action with respect to the metric yields a covariantly conserved, symmetric
energy-momentum-tensor. In particular, we may consider a Lagrangian that
only involves the first derivatives $\partial_\mu z_a$. The matrix
\bdm H_{ab}= g^{\mu\nu}\partial_\mu z_a\partial_\nu z_b\edm
represents a set of scalar fields, so that any expression of the
form\footnote{The line
element $ds^2\!=\!c^2dt^2\!-\!dx^adx^a$ corresponds to
$g_{00}\!=\!c^2,\,g^{00}\!=\!c^{-2}$. To ensure that, on flat space, the term
$\sqg\,$ reduces to unity, I set $\rule{0.03em}{0.6em}\hspace{0.1em}g
\hspace{0.1em}\rule{0.03em}{0.6em} \equiv\det(-g)/c^2$.}
\bdm {\cal
L}_\eff=\sqg\,c^2f(H)\edm yields equations of motion that do admit a
translation invariant, conserved, symmetric energy-momentum-tensor, given by
\bdm \theta^{\mu\nu}=2\,\partial^\mu z_a\partial^\nu
z_b\frac{\partial f(H)} {\partial H_{ab}}-g^{\mu\nu } f(H)\fs\edm

Free particles, for example, are characterized by
the action
\bdm S\!=\!-\sum_i m_ic\!\int \!ds_i\co\edm
where $ds_i$ is the Minkowski
line element along the world line that describes
the motion of one of these particles. For a continuous distribution,
the configuation is described by three functions $z_a(x)$ that remain
constant along the world lines. The vectors $\partial_\mu z_a(x)$ span a
three-dimensional space, orthogonal to the world line
passing through $x$. Together with the tangent vector to the
world line, $u^\mu\!=\!dx^\mu/ds$, they form a complete set of vectors, so
that the difference $dx^\mu$ between neighbouring points may be decomposed as
$dx^\mu=\partial^\mu z_a dq^a+u^\mu dq^0$. Since the vectors $u^\mu$ and
$\partial_\mu z_a$ are orthogonal, the line element takes the form
$dx^\mu dx_\mu\!=\!H_{ab}dq^adq^b+(dq^0)^2$ and the corresponding expression
for the volume element reads
$ c\sqg\;d^4\!x\!=\!\sqH\;d^4\!q$. The component $dq^0$ measures
the length in the direction of the world line, $dq^0\!=\!ds$, while the
components $dq^a$ represent the projections of $dx^\mu$ in the space
orthogonal to it, related to
\bdm dz_a\!=\!\partial_\mu z_a
dx^\mu\!=\!H_{ab}dq^b\fs\edm Denoting the mass contained in
$d^3\!z$ by $\rho_0 \,d^3\!z$, the action becomes
\bdm S=-\int\!\!\rho_0\,
d^3\!z\, c\, ds=-\rho_0 c\!\int\!\!|\!\det H| d^4\!q=-\rho_0
c^2\!\int\!\!\sqg\;
\sqH\; d^4\!x\fs\edm
This shows that the Lagrangian of a continuous distribution of free
particles is indeed of the above form, with $f(H)\!=\!-\rho_0 \sqH\,$. The
corresponding energy-momentum-tensor is given by the familiar expression
$\theta^{\mu\nu}\!=\!\rho\, u^\mu u^\nu$,
appropriate for a cloud of dust. The quantity
$\rho$
differs from the constant mass density $\rho_0$ of the ground state by a factor
that depends on the deformation: $\rho=\rho_0\sqH\,$.

The free Lagrangian may be generalized to allow for pressure. For the
energy-momentum-tensor to take the form
$\theta^{\mu\nu}\!=\!\rho\, u^\mu
u^\nu+c^{-2}(u^\mu u^\nu-g^{\mu\nu})P$, the function $f(H)$  can depend on
the matrix $H_{ab}$ only through the determinant,
\bdm{\cal L}_\eff\!=\! -\sqg\;\rho(h)c^2\co\;\;\;h=|\!\det H|\fs\edm
Accordingly, the system must exhibit invariance with respect to
volume-preserving reparametrizations,
$z_a\!\rightarrow\!\psi_a(z)$, $\det \partial \psi/\partial z\!=\!1$. The
corresponding expression for the pressure reads
\bdm P=c^2\{2 h\frac {d\rho(h)}{dh}-\rho(h)\}\fs\edm
The shape of the function $\rho(h)$ thus determines the dependence
of the pressure on the mass density and vice versa.

On flat
space, the matrix $H_{ab}$ reduces
to \bdm
H_{ab}=-\delta_{ab}+\Hbar_{ab}\co\;\;\;\Hbar_{ab}=
H_{ab}=\partial_a\xi_b+\partial_b\xi_a
-\partial_c\xi_a\partial_c\xi_b+c^{-2}\dot{\xi}_a\dot{\xi}_b\fs\edm
To account for the torsion and compression modules, it suffices to include
additional terms in the effective Lagrangian,
\bea {\cal L}_\eff\al=\al\sqg\,\sqH\;\left\{-\rho_0
c^2-\mbox{$\frac{1}{8}$}(K-\mbox{$\frac{2}{3}$}\mu)\,
\mbox{Sp}(\!\Hbar)^{2}-\mbox{$\frac{1}{4}$}\mu\,\mbox{Sp}
(\!\Hbar\rule{0em}{0.8em}^{\,2})\right.\no \al+\al
\left. L_1\,\mbox{Sp}(\!\Hbar)^3
+L_2\,\mbox{Sp}(\!\Hbar)\,\mbox{Sp}(\!\Hbar\rule{0em}{0.8em}^{\,2})+
L_3\,\mbox{Sp}(\!\Hbar\rule{0em}{0.8em}^{\,3})
+\ldots\right\}\co \nonumber\eea where
$\mbox{Sp}(\!\Hbar\rule{0em}{0.8em}^{\,2})$ stands for $H_{ab}H_{ba}$.
Note that, for the additional terms, the factor
$\sqH\,$ could just as well be dropped, as it merely amounts to a reordering:
the expansion in powers of $\Hbar_{ab}$ yields
\bea\sqH=\{\det(1\!-\!\Hbar)\}^\frac{1}{2}\al=\al
1-\mbox{$\frac{1}{2}$}\mbox{Sp}(\!\Hbar)+
\mbox{$\frac{1}{8}$}\mbox{Sp}(\!\Hbar)^2
-\mbox{$\frac{1}{4}$}\mbox{Sp}(\!\Hbar\rule{0em}{0.8em}^{\,2})\no
\al-\al\mbox{$\frac{1}{48}$}\mbox{Sp}(\!\Hbar)^3
+\mbox{$\frac{1}{8}$}\mbox{Sp}(\!\Hbar)\,
\mbox{Sp}(\!\Hbar\rule{0em}{0.8em}^{\,2 } )
-\mbox{$\frac{1}{6}$}\mbox{Sp}(\!\Hbar\rule{0em}{0.8em}^{\,3})
+\ldots\nonumber\eea

The general expression does reproduce the terms of order $\xi^2$ occurring in
${\cal
L}_2$, but the expansion necessarily also generates higher order contributions,
because the matrix $\Hbar_{ab}$ contains terms quadratic in $\xi$. Within the
covariant framework, the fact that the
effective Lagrangian must contain
terms of order $\xi^3$ is so obvious that it is barely mentioned in the
textbooks.
The reparametrization invariant Lagrangian considered above corresponds to the
special case $\mu\!=\!0,\; L_2\!=\!-\frac{1}{8}K,\; L_3\!=\!0$. At the order
of the derivative expansion considered,
these systems are characterized by two independent parameters:
$K, L_1$. In particular, the energy-momentum tensor can take the form
$\theta^{\mu\nu}\!=\!\rho\, u^\mu
u^\nu+c^{-2}(u^\mu u^\nu-g^{\mu\nu})P$ only if the torsion module vanishes,
as it is the case for a fluid.

Incidentally, the covariant formulation readily accomodates anisotropy
terms like $\Sigma_a\partial_a\xi_a\partial_a\xi_a$: the
Lagrangian relevant
for the general case is obtained by replacing the traces in the above
expression with a polynomial formed with the matrix
$\hspace{-0.1em}\Hbar_{ab}$. In the case of cubic symmetry, for instance,
there is one additional contribution of second order, proportional to $\Sigma_a
\hspace{-0.1em}\Hbar_{aa}\hspace{-0.1em}\Hbar_{aa}$. Note also that, if the
derivative expansion is carried
further, the general Lagrangian contains invariants formed with
higher derivatives of the scalar fields.

\section{General analysis at next-to-leading order}
The formulation of the effective theory discussed in the preceding
section assumes that the effective Lagrangian may be brought
to a form that is manifestly invariant with respect to the local symmetry
group. This assumption is sufficient, but not
necessary. The ferromagnet represents an interesting example, where the
effective Lagrangian fails to be manifestly invariant: under a local
symmetry transformation, ${\cal L}_\eff$ changes by a total derivative.
In principle, the same phenomenon could also take place here. The
Ward identities that express the conservation of energy and momentum
on the level of the correlation functions formed with $\theta^{\mu\nu}(x)$ only
ensure that the corresponding effective {\it action} is invariant under general
coordinate transformations. If the effective {\it Lagrangian} is invariant,
the corresponding effective action is automatically invariant, too, but
the converse is not evident. I now wish to show that the conservation laws and
the commutation relations listed in section 1 indeed imply that, with a
suitable choice of the effective field, the interaction term ${\cal
L}_3$ may be brought
to manifestly invariant form, at least to first nontrivial order in the
derivative expansion.
The technique used is brute force: I first write
down all possible terms occurring in the relevant expressions for the effective
Lagrangian and for the energy momentum tensor and then work out the
constraints imposed by the conservation laws and commutation relations.

The general expression for ${\cal L}_3$ contains terms with an odd
number of space derivatives and an even number of time derivatives.
Consider all contributions up to
to and including three derivatives. An integration by parts suffices to
eliminate terms containing the third derivative of the field. The same
operation also removes
contributions containing $\ddot{\xi}_a$. The general expression then contains
altogether 9 independent vertices: \bea {\cal L}_3\al=\al
l_1\,\dot{\xi}_a\dot{\xi}_a\partial_b\xi_b
+l_2\,\dot{\xi}_a\dot{\xi}_b\partial_a\xi_b
+ l_3\,\partial_a\xi_a\partial_b\xi_b\partial_c\xi_c\no
     \al + \al l_4\,\partial_a\xi_a\partial_b\xi_c\partial_b\xi_c
    +l_5\,\partial_a\xi_a\partial_b\xi_c\partial_c\xi_b
      +l_6\,\partial_a\xi_b\partial_a\xi_c\partial_b\xi_c\no
\al+\al
l_7\,\xi_a\xi_a\partial_b\xi_b+
l_8\,\xi_a\dot{\xi}_a\partial_b\dot{\xi}_b
+  l_9\,\xi_a\partial_b\xi_b\partial_c\partial_c\xi_a +O(p^5)\fs
\nonumber\eea
Using partial integration, all other
terms, such as $\partial_a\xi_b\partial_b\xi_c\partial_c\xi_a$, may be absorbed
in the coupling constants $l_1,\ldots\,,l_9$.

The covariant Lagrangian given in the preceding section only
contains 3 independent coupling constants at this order of the expansion:
$L_1,L_2,L_3$. I wish to show that the symmetry indeed determines all but 3
of the above 9 couplings -- except for the freedom in the choice of the
variables.
For this purpose, the effective representation of the energy-momentum-tensor
is needed to first nonleading order in $\xi$. In the case of the momentum
density,
the general expression involves 6 independent terms at this order:  \bea
\theta^{0a}_2\al=\al  p_1\,\dot{\xi}_a\partial_b\xi_b+
p_2\,\dot{\xi}_b\partial_a\xi_b+p_3\,\dot{\xi}_b\partial_b\xi_a\no
\al+\al p_4\,\xi_a\partial_b\dot{\xi}_b+
p_5\,\xi_b\partial_a\dot{\xi}_b+
p_6\,\xi_b\partial_b\dot{\xi}_a +O(p^4)\fs\nonumber\eea
Three of these may be eliminated, however, with the following observation.
The dynamical variables occurring in the framework of an effective theory
represent auxiliary quantities. In the present context, the effective field
may be subject
to the transformation \bdm \xi_a\rightarrow \xi_a
+\kappa_1\,\xi_a\partial_b\xi_b
+\kappa_2\,\xi_b\partial_a\xi_b +\kappa_3\,\xi_b\partial_b\xi_a\co\edm
without changing the content of the effective theory. Inserted in the above
expression for the momentum density, the change of variables modifies the
values of the constants occurring therein. We may exploit this
freedom and, with a suitable choice of $\kappa_1,\kappa_2,\kappa_3$, remove the
coefficients $p_4,p_5,p_6$. Without loss
of generality, the expression for the momentum density then becomes
translation invariant:
\bdm \theta^{0a}_2= p_1\,\dot{\xi}_a\partial_b\xi_b+
p_2\,\dot{\xi}_b\partial_a\xi_b+p_3\,\dot{\xi}_b\partial_b\xi_a+
O(p^4)\fs\edm

Next, I observe that energy conservation determines the energy density in terms
of $\partial_a\theta^{0a}$, up to an additive constant that represents the
energy density of the ground state. In particular, translation
invariance of $\theta^{0a}$ implies translation invariance of $\theta^{00}$.
With this simplification, the explicit expression takes the form
\bdm \theta^{00}_2=e_1\,\dot{\xi}_a\dot{\xi}_a
+e_2\,\partial_a\xi_b\partial_a\xi_b
+e_3\,\partial_a\xi_b\partial_b\xi_a
+e_4\,\partial_a\xi_a\partial_b\xi_b +O(p^4)\fs\edm
The coefficients are fixed in terms of those occurring in $\theta^{0a}$:
\bdm e_1=-\frac{\rho_0\, p_2}{2\mu}\co\;\;e_2=-\frac{p_2}{2}
\co\;\;e_3=-\frac{p_3}{2}\co\;\;e_4=-\frac{p_1}{2}\fs\edm
The energy is conserved if and only if $p_1,p_2,p_3$ obey the
condition \bdm \mu \,p_1-(K+\mbox{$\frac{1}{3}$}\mu)\,p_2+\mu \,p_3=0\fs\edm

In the case of the stress tensor,
the general expression for the first nonleading terms is rather voluminous:
eliminating second time derivatives with the equation of motion, it takes
the form
\bea
\al\al\theta^{ab}_2= s_1\,\dot{\xi}_a\dot{\xi}_b
+s_2 \,\partial_c\xi_a\partial_c\xi_b
+ s_3\,(\partial_a\xi_c\partial_c\xi_b\!+\!\partial_b\xi_c\partial_c\xi_a)
+s_4\,\partial_a\xi_c\partial_b\xi_c \no
\al+\al\! s_5\,(\partial_a\xi_b\!+\!\partial_b\xi_a)\partial_c\xi_c
+\!s_{6}\,(\xi_a\partial_c\partial_c\xi_b\!+\!\xi_b\partial_c\partial_c\xi_a)
+\!s_{7}\,\partial_c\xi\partial_a\partial_b\xi_c\no
\al+\al\! s_{8}\,\xi_c(\partial_a\partial_c\xi_b\!+\!\partial_b\partial_c\xi_a)
+\! s_{9}\,(\xi_a\partial_b\partial_c\xi_c\!+\!
          \xi_b\partial_a\partial_c\xi_c)
+\! s_{10}\,\delta_{ab}\dot{\xi}_c\dot{\xi}_c
+\! s_{11}\,\delta_{ab}\partial_c\xi_d\partial_c\xi_d\no
\al+\al\! s_{12}\,\delta_{ab}\partial_c\xi_d\partial_d\xi_c
+\! s_{13}\,\delta_{ab}\partial_c\xi_c\partial_d\xi_d
+\!s_{14}\,\delta_{ab}\xi_c\partial_d\partial_d\xi_c
+\!s_{15}\,\delta_{ab}\xi_c\partial_c\partial_d\xi_d
+O(p^4)\fs\nonumber\eea
It is advantageos to use Mathematica for the evaluation of the
derivative $\partial_a\theta^{ab}$.
The calculation shows that the conservation of momentum fixes all but two of
the coefficients $s_1,\ldots\,,s_{15}$ in terms
of those occurring in the effective Lagrangian and in the momentum density. The
freedom of choosing
two of these arbitrarily arises because we can form two independent expressions
for which the divergence $\partial_a\theta^{ab}$ vanishes identically. This
is what remains of the ambiguity
$\theta^{\mu\nu}\!\rightarrow\!\theta^{\mu\nu}
+\partial_\lambda\chi^{\lambda\mu\nu}$ mentioned earlier -- if we adopt the
above convention that leads to a translation invariant expression for
$\theta^{0a}$. Remarkably, energy and momentum conservation does not constrain
the translation invariant couplings $l_1,\ldots\,l_6$ at all:
there is a conserved and symmetric energy-momentum-tensor for any choice of
these constants. The couplings that
break translation invariance, $l_7,l_8,l_9$, on the other hand, are fixed in
terms of $p_1,p_2,p_3,l_1,\ldots\,l_6$. In particular, the conservation laws
imply that $l_7$ vanishes.

\section{Current algebra}
So far, I have only imposed the conservation laws. Now, I turn to the
requirement that the components of the energy-momentum-tensor
obey the commutation relations of current algebra:
the correlation functions of the energy-momentum-tensor only satisfy the Ward
identities if this condition is met.

At leading order of the low energy expansion, only the
tree graphs of the effective theory are relevant -- the leading term
in the low energy expansion of the
effective action is the classical action.
In the case of the commutator between two operators $A,B$,
the tree graph contributions arise from the exchange of a single
particle, generating terms with a single propagator. These are of first order
in $\hbar$ and are obtained by replacing the commutator with the Poisson
bracket: \bdm [A,B]=-i\hbar\{A,B\}\PB +O(\hbar^2)\edm
Accordingly, the current algebra conditions of section 1 are satisfied at
leading order of the low energy expansion, provided
\bdm \{P^a,\theta^{\mu\nu}(x)\}\PB=-\partial_a\theta^{\mu\nu}(x)\co\;\;\;
 \{H,\theta^{\mu\nu}(x)\}\PB=\partial_0\theta^{\mu\nu}(x)\fs\edm
To evaluate these conditions, we need to replace the time derivatives of the
field
by the canonical momentum $\pi^a=\partial {\cal L}_\eff/\partial\dot{\xi}_a$,
\bdm \pi^a=
\rho_0\, \dot{\xi}_a +2l_1\,\dot{\xi}_a\partial_b\xi_b
+l_2\,\dot{\xi}_b(\partial_a\xi_b+\partial_b\xi_a)
+l_8\,\partial_b\dot{\xi}_b\xi_a -l_8\,\partial_a\dot{\xi}_b \xi_b
-l_8\,\dot{\xi}_b\partial_a\xi_b\co\edm
and use the relations that define the Poisson bracket
\bdm \{\xi_a(x),\xi_b(x)\}\PB=\{\pi^a(x),\pi^b(x)\}\PB=0\co
\{\pi^a(x),\xi_b(x)\}\PB= \delta^a_b\delta^3(x-y)\fs\edm
For the commutator of the Hamiltonian with the energy density to
coincide with the time derivative of $\theta^{00}$, the constant $e_1$
must be related to the mass density by \bdm
e_1=\mbox{$\frac{1}{2}$}\rho_0\,c^{-2}\fs\edm
This result agrees with the standard expression
$\theta^{00}_2\!=\!\frac{1}{2}\rho_0\,\dot{\xi}_a\dot{\xi}_a +\ldots$
for the kinetic energy density.
Similarly, the bracket $\{P^a,\theta^{0b}\}\PB$ coincides with
$-\partial_a\theta^{0b}$ only if
\bdm p_1=2l_1-l_8\co\;\;\;p_2=l_2-\rho_0 \co\;\;\;p_3=l_2\fs\edm
These conditions do impose strong constraints on the coupling constants
of the Lagrangian. Together with the conservation laws, they imply that
the couplings of those terms that violate translation invariance
vanish. Moreover, they permit only three independent translation invariant
couplings.
In fact, the resulting form of the effective Lagrangian coincides with the
manifestly covariant expression given in section 4, so that all of the
coupling constants may be expressed in terms of the parameters $L_1,L_2,L_3$
occurring therein:
\bea l_1 \al=\al- \mbox{$\frac{1}{2}$}\rho_0 -\mbox{$\frac{1}{2}$}Kc^{-2}
+ \mbox{$\frac{1}{3}$}\mu c^{-2}\;,\;\;  l_2=\rho_0
-\mu c^{-2}\;,\;\;  l_3 = \mbox{$\frac{1}{2}$}K -
\mbox{$\frac{1}{3}$}\mu+8L_1 - L_3
\;,\no  l_4\al=\al  \mbox{$\frac{1}{2}$}K
+ \mbox{$\frac{1}{6}$}\mu+4L_2 \;,\;\;  l_5 =
\mbox{$\frac{1}{2}$}\mu+4L_2 + 3L_3  \;,\;\;
 l_6 =  \mu+6L_3 \co\no l_7 \al=\al 0\;,\;\; l_8 = 0\;,\;\; l_9 =
0\fs\nonumber \eea

\section{Conclusion}
This completes the explicit demonstration of the claim made above: current
algebra forces the phonons to interact, so that the wave equation for sound
necessarily contains nonlinear terms.
The Ward
identities associated with the hidden symmetry imply that, with a
suitable choice of the effective fields, the low energy structure
of the system may be described in terms of
a manifestly covariant effective Lagrangian
that involves three independent coupling
constants at the order considered. Ignoring the tiny relativistic corrections
of order $(v/c)^2$, where $v$ is the velocity of sound, the contributions from
$l_1$ and $l_2$ modify the kinetic term
according to
\bdm\mbox{$\frac{1}{2}$}\rho_0\,\dot{\xi}_a\dot{\xi}_a\rightarrow
\mbox{$\frac{1}{2}$}\rho_0\,\dot{\xi}_a\dot{\xi}_a-
\mbox{$\frac{1}{2}$}\rho_0\,\partial_b\xi_b\dot{\xi}_a\dot{\xi}_a+
\rho_0 \,\partial_a\xi_b\dot{\xi}_a\dot{\xi}_b\fs\edm
With the choice of the effective fields adopted
here, the symmetry implies that those couplings which break translation
invariance vanish:
$l_7\!=\!l_8\!=\!l_9\!=\!0$.
Moreover, the symmetry imposes one constraint among the four translation
invariant couplings $l_3,l_4,l_5,l_6$.

In the case of $\pi\pi$-scattering, Lorentz
invariance and Bose statistics imply that the leading term in the low energy
expansion of the scattering amplitude contains two constants,
$A(s,t,u)\!=\!a_1 M_\pi^2+a_2s+O(p^4)$. The symmetry
determines $a_1$ and $a_2$ in terms of a single parameter, the pion decay
constant \cite{Weinberg 1966}: $a_1\!=\!-1/F_\pi^2,\;a_2\!=\!1/F_\pi^2$. In the
present case, the leading term in the derivative expansion of the interaction
involves 9 constants $(l_1,\ldots\,,l_9)$ and the symmetry determines
these in terms of the three independent couplings ($L_1,L_2,L_3$) that occur
in the covariant expression for the effective Lagrangian. As
announced
in the introduction, the rather lengthy calculation described here merely
rederives the standard form of the effective Lagrangian and does therefore
not add anything to what is known about the physics of
the phonons. It demonstrates, however, that the same mechanism that
subjects
the Goldstone bosons of QCD to a specific interaction at low energies
is also at work in a solid, where it implies that the propagation of sound is
an intrinsically nonlinear phenomenon.

\subsection*{Acknowledgement} I am indebted to Jerzy Kijowski and Andrej
Smilga for useful remarks and discussions.

\end{document}